\documentclass[aps,prl,twocolumn,showpacs,groupedaddress]{revtex4-1} 
\usepackage{epsfig}
\usepackage{amssymb,amssymb}
\usepackage{atlasphysics}
\usepackage{subfigure}
\newlength{\figwidth}
\usepackage{longtable}
\usepackage{dcolumn}

\def\integLumi{$3.1$~pb$^{-1}$}

\begin{document}

\title{Search for Diphoton Events with Large Missing Transverse Energy
in 7~TeV Proton-Proton Collisions with the ATLAS Detector}

{CERN-PH-EP-2010-076; Submitted to PRL.}
 
\author{The ATLAS Collaboration}

\begin{abstract}
A search for diphoton events with large missing transverse energy
is presented. The data were collected with the ATLAS detector in
proton-proton collisions at $\sqrt{s} = 7$~TeV at the CERN Large Hadron Collider and
correspond to an integrated luminosity of \integLumi.
No excess of such events is observed above the Standard Model background prediction.  
In the context of a specific model with one Universal 
Extra Dimension with compactification radius $R$ and gravity-induced decays,
values of $1/R < 729$~GeV are excluded at 95\% CL, providing the most sensitive limit
on this model to date.
\end{abstract}

\maketitle

 In the Standard Model (SM), the production in proton-proton ($pp$) collisions of diphoton 
 ($\gamma \gamma$) events with large missing transverse energy (\met) is mainly due to
$W/Z + \gamma \gamma$ processes.  Taking into account the branching
ratios of $W/Z$ decays including at least one neutrino, the cross sections are
only a few femtobarns for 7~TeV $pp$ collisions. In contrast, some new physics models
predict much larger $\gamma \gamma + \met$ rates.
 This Letter reports the first $\gamma \gamma + \met$ search with LHC data, using data
 recorded with the ATLAS detector. The results are interpreted 
 in the context of a Universal Extra Dimension (UED) model. 

UED models~\cite{UEDtheory} postulate the existence of additional 
spatial dimensions in which all SM particles can propagate, leading 
to the existence for each SM particle of a series of excitations, 
known as a Kaluza-Klein (KK) tower. This analysis considers
the case of a single TeV$^{-1}$-sized UED, with compactification radius R.
The masses of the states of successive levels in the tower are separated by $\approx 1/R$.  
For a given KK level, the approximate mass degeneracy of the KK excitations is
broken by radiative corrections~\cite{UEDcutoff}.  
The lightest KK particle (LKP) is the KK photon of the first level, denoted $\gamma^*$.
At the LHC, the main UED process would be production via the strong interaction
of a pair of first-level KK quarks and/or gluons~\cite{UEDprod}, which would
decay via cascades involving other KK particles until reaching the LKP at the end of the decay chain.
If the UED model is embedded in a larger space with
$N$ additional eV$^{-1}$-sized dimensions accessible 
only to gravity~\cite{largerspace}, the LKP could decay gravitationally via
$\gamma ^* \rightarrow \gamma + G$~\cite{gravdecays}, where $G$ represents one 
of a tower of eV-spaced graviton states.  With two decay chains per event, the final state
would be $\gamma \gamma + \met + X$, where \met \  results from
the escaping gravitons and  $X$ represents SM particles emitted in the cascade decays.

The UED model considered is defined by specifying $R$ and $\Lambda $,   
 the ultraviolet cut-off used in the calculation of radiative corrections to the KK masses. This analysis treats
 $R$ as a free parameter and, following the theory calculations~\cite{UEDcutoff},  
 sets $\Lambda $ such that $\Lambda R = 20$.  For $1/R = 700$~GeV, the masses
 of the first-level KK photon, quark and gluon are $700$, $815$ and $865$~GeV,
 respectively~\cite{UEDpythia}.  The $\gamma ^*$ mass is insensitive to $\Lambda $, while other
 KK masses change by typically a few percent when varying  $\Lambda R$ in the range $10-30$.
 The gravitational decay widths of the KK particles are set by $N$ and 
 $M_D$, the Planck scale in the ($4+N$)-dimensional theory.   For the chosen values of 
 $N = 6$ and $M_D = 5$~TeV, and provided $1/R < 1$~TeV, the  LKP is the only KK particle to have
 an appreciable rate of gravitational decay.
The same parameter values were used in the only previous study of this model, in which the 
D0 experiment excluded at 95\% CL values of $1/R < 477$~GeV~\cite{DZERO}.
 
 Monte Carlo (MC) signal samples were produced for a range of $1/R$ values using the  
 implementation~\cite{UEDpythia} of the UED model in PYTHIA~\cite{pythia} version 6.421, and
 using the MC09 parameter tune~\cite{ATLAStune}. The MC samples were processed through
 the ATLAS detector simulation~\cite{ATLASsim} based on GEANT4~\cite{geant4}.  In addition to the two 
 high transverse energy (\ET) photons and
 large \met, the signal events typically include several high-\ET \ jets due to the cascade decays, 
 with the \ET \ spectrum of the leading jet peaking at
 $\approx 100$~GeV for $1/R = 700$~GeV.
  
The ATLAS detector~\cite{DetectorPaper} is a multipurpose particle
physics apparatus with a forward-backward symmetric cylindrical
geometry and nearly 4$\pi$ solid angle coverage.
ATLAS uses a Cartesian right-handed coordinate system, with
the nominal collision point at the origin.  The anticlockwise beam
direction defines the positive $z$-axis, while the positive $x$-axis
points from the collision point to the centre of the LHC ring and the
positive $y$-axis points upward.  The angles $\phi$ and $\theta$\ are
the azimuthal and polar angles.  The pseudorapidity is defined as
$\eta = -\ln\left[\tan\left(\theta / 2\right)\right]$.
Closest to the beamline are tracking detectors which use layers of
silicon-based and straw-tube detectors,  located inside a thin superconducting solenoid 
that provides a 2~T magnetic field, to measure the trajectories of
charged particles. The solenoid is surrounded by
a hermetic calorimeter system.  A liquid-argon (LAr) sampling calorimeter is divided into
a central barrel calorimeter and two end-cap calorimeters, each housed in a
separate cryostat. Fine-grained LAr electromagnetic (EM) calorimeters,
with excellent energy resolution, 
provide coverage for $\left|\eta\right| < 3.2$. In the region
$\left|\eta\right| < 2.5$, the EM calorimeters are segmented into three
longitudinal layers and the second layer, in which 
most of the EM shower energy is deposited,
is divided into cells of  granularity of $\Delta \eta \times
\Delta \phi = 0.025 \times 0.025$.
A presampler, covering $\left|\eta\right| < 1.8$,  is used to correct for
energy lost upstream of the calorimeter. An iron-scintillator
tile calorimeter provides hadronic coverage in the range 
$\left|\eta\right| < 1.7$. In the end-caps
($\left|\eta\right| > 1.5$), LAr hadronic calorimeters match the outer $\left|\eta\right|$
limits of the end-cap EM calorimeters.
LAr forward calorimeters provide both EM
and hadronic energy measurements, and extend the coverage to $\left|\eta\right| < 4.9$.
Outside the calorimeters is an extensive
muon system including large superconducting toroidal magnets.

The reconstruction of photons is described in detail in Ref.~\cite{directphoton}. To
select photon candidates, EM calorimeter 
clusters were required to pass several quality criteria and
to lie outside problematic calorimeter regions. 
Photon candidates were required to have $\left|\eta\right| < 1.81$ and to
be outside the transition region $1.37 < \left|\eta\right| < 1.52$
between the barrel and the end-cap calorimeters.
The analysis uses a ``loose'' photon selection,
which includes cuts on the energy in the hadronic calorimeter as well as on
variables that require the transverse width of the shower, measured in the second 
EM calorimeter layer, be consistent with the narrow width expected for an EM shower. 
The loose selection provides a high photon efficiency with modest rejection against
the background from jets. 

The reconstruction of \met \ is based on topological calorimeter clusters~\cite{WZpaper} with $|\eta | < 4.5$
that are seeded by any cell with energy higher than four times its noise level. In an iterative procedure, 
the cluster grows by including all neighbouring cells with energy 
higher than twice the noise, plus all cells neighbouring the boundary of this 
three-dimensional collection. Each cluster is classified as
EM or hadronic, depending on its topology, and the cluster energy is calibrated to
correct for the non-compensating calorimeter response, energy losses in 
dead material, and out-of-cluster energies.
Events reconstructed with large \met \  were studied in detail 
with early data~\cite{jetcleaning}. Rare background events with large transverse energies, 
unrelated to the collision and concentrated in a few cells, due mainly to discharges 
and noise, have been observed. Cuts were applied to eliminate such backgrounds,
rejecting less than 0.05\% of the selected events while having a negligible impact on the signal efficiency.

The data sample was collected during stable beam periods of 7~TeV $pp$ collisions
at the LHC,
and corresponds to an integrated luminosity of \integLumi.
The events selected had to satisfy a trigger requiring at least one loose photon candidate
with $\ET > 20$~GeV, and had to contain at least one reconstructed primary vertex
consistent with the average beam spot position and with at least three associated tracks.
The trigger and vertex requirements are $\approx 99$\% efficient for signal MC events.
The presence of multiple $pp$ collisions within the same bunch crossing,
known as ``pileup'', can be analysed by examining $N_{vtx}$, the number of
reconstructed primary vertices in each event. 
In this data sample, the average value of $N_{vtx}$ was $\approx 2.1$.
The MC signal samples included the simulation of pileup and
were weighted to match the $N_{vtx}$ distribution observed in data.

\begin{figure}[hbpt]
\begin{center}
\includegraphics[width=0.48\textwidth]{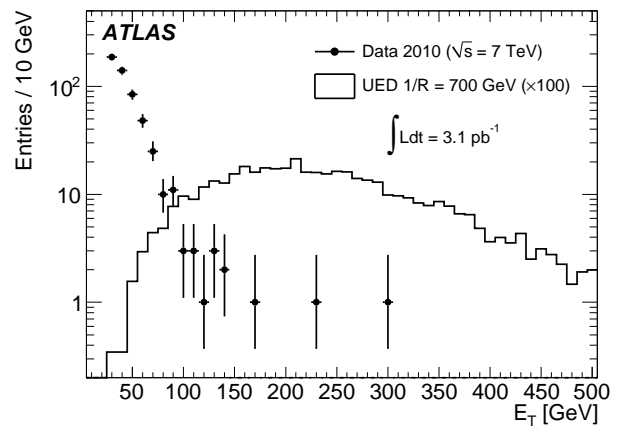}
\end{center}
\caption{\ET \ spectrum of the leading photon for the $\gamma \gamma$ 
candidate sample and for UED $1/R = 700$~GeV MC events
(normalized to 100 times the leading order (LO) cross section).}
\label{fig:ETgamma}
\end{figure}

Events were retained if they had at least 
two photon candidates, each with $\ET > 25$~GeV. In addition,
a photon isolation cut was applied, wherein
the \ET \  in a radius of $0.2$ in the $\eta$-$\phi$ space around the 
center of the cluster, excluding the cells belonging to the cluster in a
region corresponding to $5 \times 7$ cells in $\eta\times\phi$ in the second layer of
the EM calorimeter, had to be less than 35~GeV.
This requirement had a signal efficiency greater than $95$\% 
but rejected some of the background from multijet events. 
An event in which each of the two 
photon candidates satisfied the loose photon cuts was
considered a $\gamma \gamma $ candidate event.
An independent ``misidentified jet'' control sample, enriched in 
events with jets misidentified as photons,  was defined as those events 
where at least one of the photon candidates did not pass the loose photon identification.
After all cuts, the $\gamma \gamma$  and misidentified jet samples totaled 520 and
7323 events, respectively.
Figure~\ref{fig:ETgamma} shows the \ET \ spectrum of the leading photon
for the $\gamma \gamma $ candidates and for UED $1/R = 700$~GeV MC events; 
the UED spectrum extends to much higher \ET \ values.

\begin{figure}[hbpt]
\begin{center}
\includegraphics[width=0.48\textwidth]{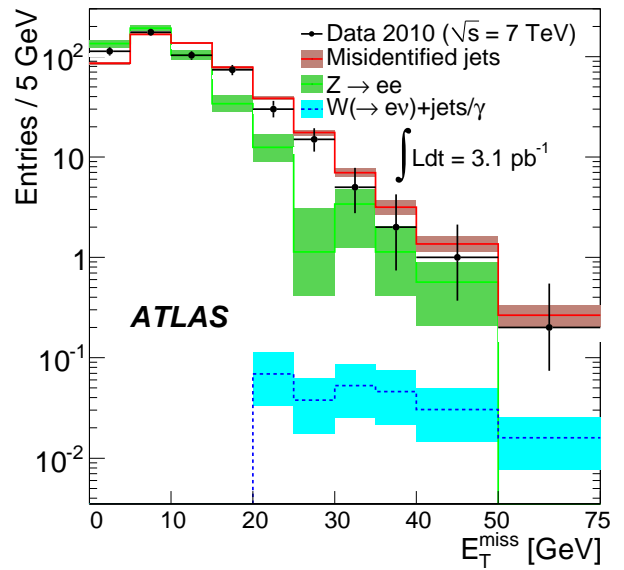}
\end{center}
\caption{\met \ spectra for the $\gamma \gamma$ candidates, for
the \Zee \ and misidentified jet samples used to model the QCD background
(each normalized to the number of $\gamma \gamma$ candidates
with $\met < 20$~GeV), and for  the $W(\rightarrow e \nu)$ + jets/$\gamma$ background (normalized to its
expected total of $\approx 0.4$ events).
Variable sized bins are used, and the vertical error bars and 
shaded bands show the statistical errors.}
\label{fig:bkgnd}
\end{figure}

The background was evaluated entirely using data.
Non-collision backgrounds, such as cosmic rays and beam-halo
events, are reduced to a negligible level by the selection cuts.
The main background source, referred to hereafter as QCD background,
 arises from a mixture of SM processes  including $\gamma \gamma$ production, and 
$\gamma $ + jet and multijet events with at least one jet misidentified as a photon.
 With the loose photon identification, it is expected that $\gamma $ + jet and multijet events
 dominate, with only a small $\gamma \gamma$ contribution. 
The misidentified jet sample provided a model of the
\met \ response for events with jets faking photons. 
The response for $\gamma \gamma$ events was modeled using
the \met \ spectrum measured in a high purity sample of \Zee \ events, selected by a combination of
kinematic cuts and electron identification requirements~\cite{WZpaper}.
The \met \ spectrum for \Zee \ events, which is  
dominated by the calorimeter response to two genuine EM objects,
was verified in MC to model the \met \ response in
SM $\gamma \gamma $ processes, despite their kinematic differences.
As shown in Fig.~\ref{fig:bkgnd}, \Zee \ events typically have somewhat lower \met \ values than events of the misidentified jet
sample, as expected since the presence of jets faking photons should result in a broader \met \ distribution. 
 The spectrum for the $\gamma \gamma $ candidates, which for low \met \
 is dominated by the QCD background with an unknown mixture
 of events with zero, one, and two fake photons, lies between these two samples.  The
\met \ spectrum of the total QCD background was modeled by a weighted sum of the spectra of the
 \Zee \ and misidentified jet samples.   The QCD background was normalized to have 
 the same number of events  as the $\gamma \gamma $ candidate sample in the 
 region $ \met < 20$~GeV, where any UED signal contribution can be neglected.
The relative contributions of the \Zee \ and misidentified 
jet samples were determined by fitting the QCD background shape to the
 \met \ spectrum of the $\gamma \gamma$  candidates in this same low \met \ region.
The fraction attributed to $\gamma \gamma$ production, as modeled with the \Zee \ distribution,
was determined to be $(36 \pm 22$)\%. The search result is not very sensitive to the exact
composition of the QCD background, and 
the fit error was used to determine systematic uncertainties on the background prediction. 

A small additional background results from \Wen \ events, which have genuine \met \ 
  and which can pass the selection if the electron is misidentified as a
photon and the second photon is either a real photon in $W\gamma$ events
or 	a jet faking a photon in $W$ + jets events.  A high purity sample
of inclusive \Wen \  events was selected by a combination of kinematic and electron
identification cuts~\cite{WZpaper}.
Requiring in addition a loose photon with $\ET^\gamma > 25$~GeV,
a ``$W+\gamma $'' sample of only 5 events was selected.
Accounting for the probability for an  electron to be misidentified as a loose photon, 
as determined using the \Zee \ sample, the total background contribution due to \Wen \  events
was then estimated to be only $\approx 0.4$ events.   Since the
number of $W\gamma$ events was too small to measure their \met \ spectrum,
a sample of $W$ + jets events was used instead, requiring
a jet reconstructed with an anti-\kT \ clustering
algorithm~\cite{antikt} with radius parameter $0.4$ and $\ET^j > 25$~GeV.
The $W(\rightarrow e \nu)$ + jets/$\gamma$ background contribution was
then estimated by normalizing the $W$ + jets \met \ spectrum
to the expected total of $\approx 0.4$ events, as shown on Fig.~\ref{fig:bkgnd}.

\begin{figure}[hbpt]
\begin{center}
\includegraphics[width=0.48\textwidth]{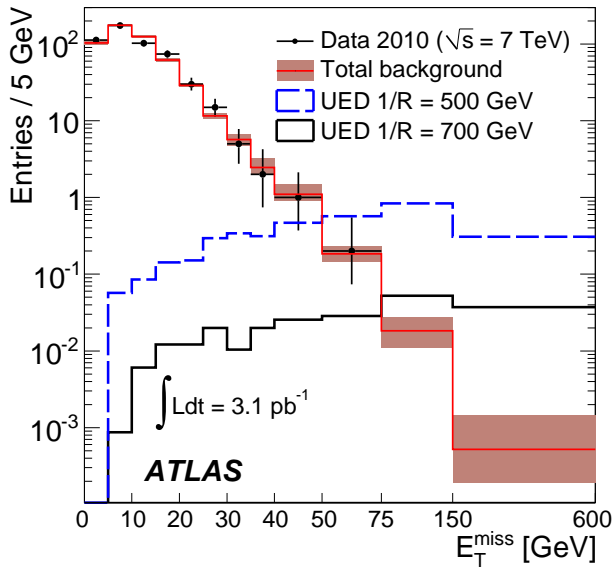}
\end{center}
\caption{\met \ spectrum for the $\gamma \gamma$ candidates, 
compared to the total SM background as estimated from data. Also shown
are the expected UED signals for $1/R = 500$~GeV and $700$~GeV.
Variable sized bins are used, and the vertical error bars and 
shaded bands show the statistical errors.}
\label{fig:met}
\end{figure}

Figure~\ref{fig:met} shows the \met \ spectrum of the $\gamma \gamma$
candidates, superimposed on the total background prediction, as well as example UED signals.
Table~\ref{tab:events} summarizes the number of observed $\gamma \gamma$ candidates, as
well as the expected backgrounds and example UED signal contributions, in several \met \ ranges.
The QCD background dominates, and falls steeply with rising \met, while the \Wen \ background is
very small, and flatter as a function of \met. The UED signals would peak at large values of \met.
There is good agreement between the data
and predicted background over the entire \met \ range, with no indication
of an excess at high \met \ values.

\begin{table*}[hbpt]
\caption{The number of observed $\gamma \gamma$ candidates, as well as the
SM backgrounds estimated from data and expected UED signal for $1/R$ values of $500$ and $700$~GeV,
given in various \met \ ranges. The uncertainties are statistical only.
 The first row, for $\met < 20$~GeV, is the control region used to normalize the 
QCD background  to the number of observed $\gamma \gamma$ candidates.}
\label{tab:events}
\begin{ruledtabular}
\begin{tabular}{ccccccc} 
\met \ range  & Data    & \multicolumn{3}{c}{Predicted background events} & \multicolumn{2}{c}{Expected UED signal events} \\
   (GeV)         & events & Total & QCD & $W(\rightarrow e \nu)$ + jets/$\gamma$  & $1/R = 500$ GeV & $1/R = 700$ GeV \\ \hline
  \hspace{0.06in}0 - 20  & 465 & $465.0\pm 9.1$ & $465.0 \pm 9.1$ &            -    & $0.28\pm 0.06$ & $0.02\pm 0.01$  \\ \hline
20 - 30  & \hspace{0.06in}45 & $\hspace{0.06in}40.5\pm 2.2$ & $\hspace{0.06in}40.41 \pm 2.17$ & $0.11\pm 0.07 $ & $0.45\pm 0.07$ & $0.03\pm 0.01$  \\
30 - 50  & \hspace{0.12in}9 & $\hspace{0.06in}10.3\pm 1.3$ & $\hspace{0.06in}10.13 \pm 1.30$ & $0.16\pm 0.10$ & $1.60\pm 0.12$ & $0.08\pm 0.01$ \\
50 - 75  & \hspace{0.12in}1 & $\hspace{0.12in}0.93\pm 0.23$ & $\hspace{0.12in}0.85 \pm 0.23$ &  $0.08\pm 0.05$ &  $2.84\pm 0.16$ & $0.14\pm 0.01$ \\
\hspace{0.12in}$>$ 75  & \hspace{0.12in}0 & \hspace{0.12in}$0.32\pm 0.16$ & $\hspace{0.12in}0.28 \pm 0.15$ & $0.04\pm 0.03$ 
                                																	& $\hspace{-0.06in}40.45\pm 0.62$ &  $4.21\pm 0.06$ \\
\end{tabular}
\end{ruledtabular}
\end{table*}

The signal search region was chosen to be $\met > 75$~GeV, before looking at the data,
to obtain the best sensitivity to the UED signal.
 In the signal region, there are zero observed events, compared to an 
expectation of $ 0.32 \pm 0.16  (stat.)^{+0.37}_{-0.10} (syst.)$ background events.
The systematic uncertainty was derived by
studying variations of the background determination, including
varying within its error the $\gamma \gamma$ fraction determined in the fit of 
the QCD background, varying the definition of the misidentified jet sample, 
and eliminating the photon isolation cut.

The UED signal efficiency, determined from MC, 
 increases smoothly from $\approx 43$\% for $1/R = 500$~GeV to $\approx 48$\% for $1/R = 700$~GeV, with the
 lower efficiencies for smaller $1/R$ due mostly to the $\met > 75$~GeV  definition of the signal region. 
The various relative systematic uncertainties on the extraction of the UED signal cross section are summarized in
Table~\ref{tab:systematics}, including the dominant 11\% uncertainty on the integrated luminosity~\cite{lumi}. 
Uncertainties on the efficiency 
for reconstructing and identifying the $\gamma \gamma$ pair arise mainly due to differences between
MC and data in the distributions of the photon identification variables,
the need to extrapolate to the higher \ET \ values (see Fig.~\ref{fig:ETgamma})
typical of the UED photons, the impact of the photon quality cuts,  varying the scale of the photon \ET \ cut,
and uncertainties in the detailed material composition of the detector. Together these provide a
systematic uncertainty of 4\%.  The influence of pileup, evaluated by comparing MC samples
             with and without pileup, gives a systematic uncertainty of 2\%. Systematic effects on the \met \ 
reconstruction~\cite{WZpaper}, including pileup,  varying the cluster energies
within the current uncertainties, and varying the expected \met \ resolution 
between the measured performance and MC expectations, combine to
give a 1\% uncertainty on the signal efficiency.
Finally, the 1\% statistical error on the signal efficiency as determined by MC is treated as
a systematic uncertainty on the result. 
Adding in quadrature, the total systematic uncertainty on the signal yield
is 12\%.

\begin{table}[hbpt]
\caption{Relative systematic uncertainties on the expected
UED signal yield. For more details, see the text.}
\label{tab:systematics}
\begin{ruledtabular}
\begin{tabular}{lc} 
 Source of uncertainty & Uncertainty  \\ \hline
 Integrated luminosity & $\hspace{0.02in}11$\%        \\
 Photon reconstruction and identification & $\hspace{0.08in}4$\% \\
 Effect of pileup & $\hspace{0.08in}2$\% \\
 \met \ reconstruction and scale & $\hspace{0.08in}1$\%  \\ 
Signal MC statistics & $\hspace{0.08in}1$\%  \\ \hline
 Total &  $\hspace{0.02in}12$\%                                    \\
 \end{tabular}
\end{ruledtabular}
\end{table}

\begin{figure}[hbpt]
\begin{center}
\includegraphics[width=0.48\textwidth]{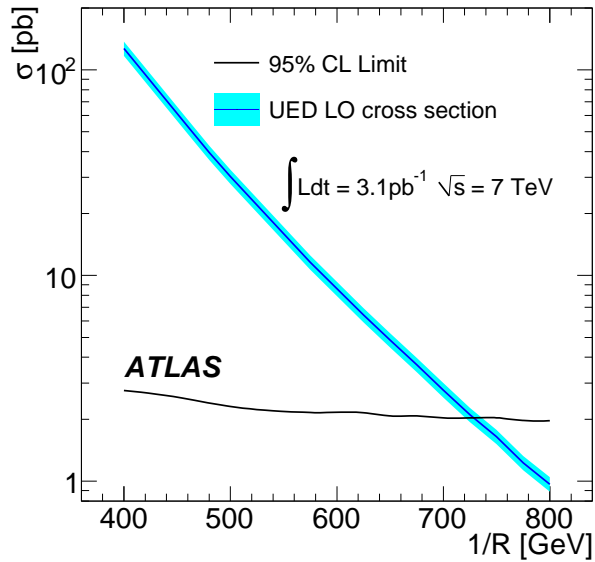}
\end{center}
\caption{95\% CL upper limits on the UED production cross section, and the LO
theory cross section prediction, as a function of $1/R$. The shaded band shows
the PDF uncertainty.}
\label{fig:limit}
\end{figure}

Given the good agreement between the measured \met \ spectrum and the expected 
background, a limit was set on $1/R$ in the specific UED model considered here.
A Bayesian approach was used to calculate a limit based on the number
of observed and expected events with $\met > 75$~GeV.
A Poisson distribution was used as the likelihood function for the expected
number of signal events, and a flat prior was used for the signal cross section. Log-normal priors were used
for the various sources of uncertainty, which were treated as nuisance parameters.
It was verified that the result is not very sensitive to the detailed form of the assumed priors.
Figure~\ref{fig:limit} depicts the resulting
95\% CL upper limit within the context of the UED model considered, together with the LO UED cross section 
as a function of $1/R$. The LO cross section was used since higher order corrections have not
been calculated for the UED model.  An uncertainty on the signal cross section due to 
parton distribution functions (PDF)  was determined by comparing the predictions using MRST2007~\cite{mrst2007}
PDF's with those from the full set of error PDF's of CTEQ6.6~\cite{cteq66}. The resultant uncertainty,
namely $\pm 8$\% essentially independent of $1/R$, is shown by the width of the theory curve band.
The observed 95\% CL exclusion region is $1/R < 729$~GeV. The result depends weakly on the
systematic uncertainties, and would only increase
to $732$~GeV if they were neglected. Changing the \met \ cut to
$60$ or $90$~GeV would change the limit by only a few GeV. A cross check using a higher purity
$\gamma \gamma$ sample, achieved by requiring that both photons pass tighter identification cuts that 
reject more of the background from jets, produced a consistent result.

In conclusion, a search for $\gamma \gamma$ events with large \met, conducted using a
\integLumi~sample of 7~TeV $pp$ collisions recorded with the ATLAS detector at the LHC, 
found  no evidence of an excess
above the SM prediction.  The results were used to set limits on a specific model with one 
UED and gravity-induced LKP decays, excluding at the 95\% CL values of $1/R <  729$~GeV, and 
significantly surpassing the only existing experimental limit~\cite{DZERO} on this model. 

We wish to thank CERN for the efficient commissioning and operation of the LHC during this initial high-energy 
data-taking period as well as the support staff from our institutions without whom ATLAS could not be operated efficiently.

We acknowledge the support of ANPCyT, Argentina; YerPhI, Armenia; ARC, Australia; BMWF, Austria; ANAS, 
Azerbaijan; SSTC, Belarus; CNPq and FAPESP, Brazil; NSERC, NRC and CFI, Canada; CERN; CONICYT, 
Chile; CAS, MOST and NSFC, China; COLCIENCIAS, Colombia; MSMT CR, MPO CR and VSC CR, Czech Republic; 
DNRF, DNSRC and Lundbeck Foundation, Denmark; ARTEMIS, European Union; IN2P3-CNRS, CEA-DSM/IRFU, 
France; GNAS, Georgia; BMBF, DFG, HGF, MPG and AvH Foundation, Germany; GSRT, Greece; ISF, MINERVA, 
GIF, DIP and Benoziyo Center, Israel; INFN, Italy; MEXT and JSPS, Japan; CNRST, Morocco; FOM and NWO, 
Netherlands; RCN, Norway;  MNiSW, Poland; GRICES and FCT, Portugal;  MERYS (MECTS), Romania;  
MES of Russia and ROSATOM, Russian Federation; JINR; MSTD, Serbia; MSSR, Slovakia; ARRS and MVZT, 
Slovenia; DST/NRF, South Africa; MICINN, Spain; SRC and Wallenberg Foundation, Sweden; SER,  SNSF and 
Cantons of Bern and Geneva, Switzerland;  NSC, Taiwan; TAEK, Turkey; STFC, the Royal Society and 
Leverhulme Trust, United Kingdom; DOE and NSF, United States of America.  

The crucial computing support from all WLCG partners is acknowledged gratefully, in particular from CERN and 
the ATLAS Tier-1 facilities at TRIUMF (Canada), NDGF (Denmark, Norway, Sweden), CC-IN2P3 (France), 
KIT/GridKA (Germany), INFN-CNAF (Italy), NL-T1 (Netherlands), PIC (Spain), ASGC (Taiwan), RAL (UK) and 
BNL (USA) and in the Tier-2 facilities worldwide.

\newpage

\begin{widetext}
\begin{flushleft}
{\Large The ATLAS Collaboration}

\bigskip

G.~Aad$^{\rm 48}$,
B.~Abbott$^{\rm 111}$,
J.~Abdallah$^{\rm 11}$,
A.A.~Abdelalim$^{\rm 49}$,
A.~Abdesselam$^{\rm 118}$,
O.~Abdinov$^{\rm 10}$,
B.~Abi$^{\rm 112}$,
M.~Abolins$^{\rm 88}$,
H.~Abramowicz$^{\rm 153}$,
H.~Abreu$^{\rm 115}$
% ...
\bigskip

% ...
$^{10}$ Institute of Physics, Azerbaijan Academy of Sciences, H. Javid Avenue 33, AZ 143 Baku, Azerbaijan\\
$^{11}$ Institut de F\'isica d'Altes Energies, IFAE, Edifici Cn, Universitat Aut\`onoma  de Barcelona,  ES - 08193 Bellaterra (Barcelona), Spain\\
% ...
$^{48}$ Albert-Ludwigs-Universit\"{a}t, Fakult\"{a}t f\"{u}r Mathematik und Physik, Hermann-Herder Str. 3, D - 79104 Freiburg i.Br., Germany\\
$^{49}$ Universit\'e de Gen\`eve, Section de Physique, 24 rue Ernest Ansermet, CH - 1211 Geneve 4, Switzerland\\
% ...
$^{88}$ Michigan State University, Department of Physics and Astronomy, High Energy Physics Group, East Lansing, MI 48824-2320, United States of America\\
% ...
$^{111}$ University of Oklahoma, Homer L. Dodge Department of Physics and Astronomy, 440 West Brooks, Room 100, Norman, OK 73019-0225, United States of America\\
$^{112}$ Oklahoma State University, Department of Physics, 145 Physical Sciences Building, Stillwater, OK 74078-3072, United States of America\\
% ...
$^{115}$ LAL, Univ. Paris-Sud, IN2P3/CNRS, Orsay, France\\
% ...
$^{118}$ Oxford University, Department of Physics, Denys Wilkinson Building, Keble Road, Oxford OX1 3RH, United Kingdom\\
% ...
$^{153}$ Tel Aviv University, Raymond and Beverly Sackler School of Physics and Astronomy, Ramat Aviv, IL - Tel Aviv 69978, Israel\\
% ...
\end{flushleft}

\end{widetext}

\end{document}